*Article*

# Angular Position Sensor Based on Anisotropic Magnetoresistive and Anomalous Nernst Effect

**Jiaqi Wang** [1,†], **Hang Xie** [1,†] **and Yihong Wu** [1,2,*]

[1] Department of Electrical and Computer Engineering, National University of Singapore, Singapore 117583, Singapore
[2] National University of Singapore (Chong Qing) Research Institute, Chongqing 401123, China
[*] Correspondence: elewuyh@nus.edu.sg
[†] These authors contributed equally to this work.

**Abstract:** Magnetic position sensors find extensive applications in various industrial sectors and consumer products. However, measuring angles in the full range of 0° - 360° in a wide field range using a single magnetic sensor remains a challenge. Here, we propose a magnetic position sensor based on a single Wheatstone bridge structure made from a single ferromagnetic layer. By measuring the anisotropic magnetoresistance (AMR) signal from the bridge and two sets of anomalous Nernst effect (ANE) signals from the transverse ports on two perpendicular Wheatstone bridge arms concurrently, we show that it is possible to achieve 0° - 360° angle detection using a single bridge sensor. The combined use of AMR and ANE signals allows to achieve a mean angle error in the range of 0.51° - 1.05° within a field range of 100 Oe – 10,000 Oe.

## 1. Introduction

Precise measurements of angular positions are indispensable across diverse domains, including manufacturing, space exploration, internet of things (IoT), medical technology, consumer products, navigation, industrial automation, etc. [1-7]. Commercial magnetic angle sensing device typically employs Hall effect [8], anisotropic magnetoresistance (AMR) [9], giant magnetoresistance (GMR) [10], and tunnel magnetoresistance (TMR) sensors [11]. To achieve full 360° detection, it is common to employ two orthogonally positioned Hall effect, GMR, or TMR sensors to obtain the sine and cosine signals from which the angle position can be derived using an arctangent function. Although AMR sensors typically offer superior signal-to-noise ratio and high accuracy compared to the Hall effect, GMR and TMR sensors, two AMR sensors can only detect the angle in the range of 0° - 180°, and an additional sensor is required to achieve full 360° detection [12,13]. This would increase the complexity of sensor design and manufacturing cost. Furthermore, the GMR and TMR sensors have a limited dynamic range (usually less than 1,000 Oe), which poses limitations in some practical applications which require a large field range.

Recently, several angular position sensors based on emerging physical phenomena such as spin Hall magnetoresistance (SMR) and spin orbit torque (SOT) have been demonstrated. These sensors typically have a much simpler design as compared to the conventional AMR, GMR and TMR sensors. In fact, all these sensors, including the SMR sensor [14], spin torque gate (STG) sensor [15], SOT-enabled anomalous Hall (AHE) vector magnetometer [16], and SOT-based magnetic angular sensor [17] are all based merely on a simple ultrathin ferromagnet (FM)/heavy metal (HM) bilayer without any magnetic bias, which greatly simplifies the sensor design and reduces the manufacturing cost. However, the SOT-based sensors typically have a relatively small dynamic range which may limit their applications in settings with large environmental magnetic field. Alternatively, the unidirectional magnetoresistance (UMR) can also be explored for 0° - 360° angle detection [18], but it faces significant challenges and limitations due to the substantial noise in the UMR signals, which adversely impacts the performance and accuracy of the sensor. Here, we propose an extremely simple angular position sensor based on the AMR and anomalous Nernst effect (ANE) in a single ferromagnetic (FM) layer, which allows for 0° - 360° angle detection using a single Wheatstone bridge. Compared to the UMR signals, the ANE signals exhibit low noise and nearly zero offset, and importantly, it is readily accessible in almost all conductive ferromagnetic films provided there is a vertical temperature gradient. The latter is conveniently achieved through the sending current. We demonstrate experimentally that, by simultaneously measuring the first harmonic longitudinal and second harmonic transverse voltages of a CoFeB-based bridge device under an ac sensing current, it is possible to detect angles within 0° - 360° at a mean angle error of 0.51° - 1.05° with a dynamic field range of 100 Oe – 10,000 Oe.

## 2. Experimental details

A stack consisting of CoFeB (3, 6, 10, 15 nm)/MgO (2 nm)/Ta (1.5 nm) from bottom to top is deposited on the Si/SiO$_2$ substrate by magnetron sputtering with a base pressure of $2 \times 10^{-8}$ Torr and a working pressure of $3 \times 10^{-3}$ Torr, respectively. The Microtech LaserWriter system, equipped with a 405 nm laser, is employed for patterning the device into Wheatstone bridge. Each Wheatstone bridge arm has a dimension of 30 μm × 200 μm. Following the deposition of the film stack and patterning of the device, electrodes and contact pads consisting of Ta (5 nm)/Cu (100 nm)/Pt (10 nm) are formed on the four terminals of the bridge for electrical transport measurements. Finally, the devices are thermally annealed at 250 °C for 1 hour in a vacuum furnace with a pressure of less than $1 \times 10^{-5}$ Torr. For electrical measurements, we use a Keithley 6221 current source to supply an ac current to two diagonal terminals of the bridge, and an MFLI lock-in amplifier from Zurich Instruments is used to measure both the first harmonic bridge output and second harmonic Hall voltage signals from the two adjacent arms. All the electrical measurements are performed in a Quantum Design Versalab PPMS system with angle rotator.

## 3. Results

*3.1. Derivation of first and harmonic signals*

Figure 1a,b show the layer structure of the sensor and schematic of the measurement setup, respectively, where $\vec{H}_{ext} = H_{ext}(\sin\theta_H \cos\varphi_H, \sin\theta_H \sin\varphi_H, \cos\theta_H)$ is an in-plane external field with azimuthal angle $\varphi_H$ and polar angle $\theta_H$ ($\theta_H = 90°$ in the present case). At large $H_{ext}$, the magnetization ($\vec{M}$) of the FM layer will be aligned with that of the external field, and therefore, its direction is: $\hat{m} = \frac{\vec{M}}{|\vec{M}|} = (\sin\theta_H \cos\varphi_H, \sin\theta_H \sin\varphi_H, \cos\theta_H)$. When a charge current (I) flows through the bridge arms consisting of CoFeB, a temperature gradient $\nabla T$ will be established in the film thickness direction (i.e., z-direction) due to the difference in thermal conductivity between air and the substrate [19], as shown in Figure 1a. The magnitude of $\nabla T$ is proportional to the power dissipation, i.e., $|\nabla T| \propto I^2 R$, where R is the resistance of the device [20]. This in turn will induce a transverse voltage signal in the respective arms due to the ANE in CoFeB. Phenomenologically, the ANE-induced electric field can be written as $\vec{E} = \frac{S_{ANE}}{|\vec{M}|} \vec{M} \times \nabla T$, where $S_{ANE}$ is the anomalous Nernst effect coefficient, indicating the strength and sign of the ANE for a particular material [21].

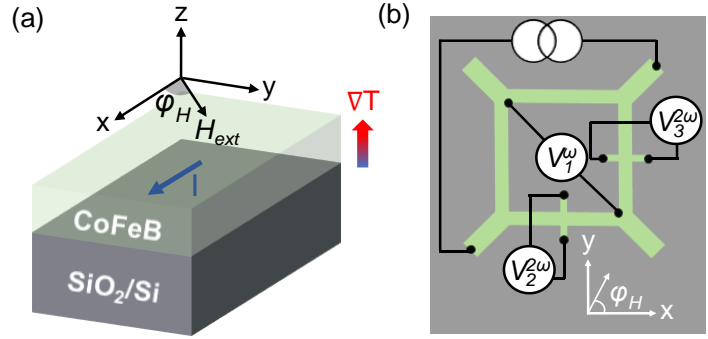

**Figure 1.** (**a**) Illustration of the layer structure of the angle sensing device; (**b**) Schematic of the harmonic measurement setup.

Without loss of generality, the total electric field ($\vec{E}$) in FM layer can be expressed in a generalized Ohm's law as follows [20,22]:

$$\vec{E} = \rho_0\vec{J} + \frac{\Delta\rho_{OMR}}{\left|\vec{H}_{ext}\right|^2}(\vec{J}\cdot\vec{H}_{ext})\vec{H}_{ext} + \frac{\Delta\rho_{AMR}}{\left|\vec{M}\right|^2}(\vec{J}\cdot\vec{M})\vec{M} - \frac{\rho_{OHE}}{\left|\vec{H}_{ext}\right|}\vec{J}\times\vec{H}_{ext} - \frac{\rho_{AHE}}{\left|\vec{M}\right|}\vec{J}\times\vec{M} \\ + \frac{S_{ANE}}{\left|\vec{M}\right|}\vec{M}\times\nabla T, \quad (1)$$

where $\rho_0$ is the longitudinal resistivity, $\vec{J}$ is the current density. $\Delta\rho_{OMR}$, $\Delta\rho_{AMR}$, $\rho_{OHE}$, and $\rho_{AHE}$ denote the resistivity changes due to the ordinary magnetoresistance (OMR), AMR, ordinary Hall effect (OHE) and AHE, respectively. The sixth term is the contribution from the ANE. As mentioned above, $\nabla T$ can be expressed as: $\nabla T = KI^2(0, 0, 1)$, where K is a device-specific coefficient, indicating a proportional relationship between $|\nabla T|$ and $I^2$. As shown in Figure 1b, three voltage signals, namely $V_1^\omega$, $V_2^{2\omega}$, and $V_3^{2\omega}$, were acquired simultaneously with an ac current ($I = I_0 \sin\omega t$) applied to the two current terminals. $V_1^\omega$ is the first harmonic signal induced by the AMR effect, whereas $V_2^{2\omega}$ and $V_3^{2\omega}$ are the second harmonic signals caused by the ANE in the two adjacent arms, named as arm-x and arm-y, respectively. The longitudinal resistance of arm-x and arm-y can be expressed as $R_0 = \frac{L\rho_0}{A}$, where A and L are the cross-sectional area and length of the arm, respectively. The current density in the two arms can be expressed as $\vec{J}_x = \frac{I_0 \sin\omega t}{2A}(1,0,0)$, $\vec{J}_y = \frac{I_0 \sin\omega t}{2A}(0,1,0)$, respectively. By substituting the expressions of $\nabla T$, $\vec{H}_{ext}$, $\vec{J}_x$, and $\hat{m}$ into Equation (1), the longitudinal voltage of arm-x ($V_{arm-x}$) can be expressed as:

$$V_{arm-x} = L\vec{E}\cdot\hat{x} = \left(\frac{L\rho_0 I_0}{2A} + \frac{L\Delta\rho_{OMR}I_0}{2A}\sin^2\theta_H\frac{1+\cos 2\varphi_H}{2}\right.\\ \left. + \frac{L\Delta\rho_{AMR}I_0}{2A}\sin^2\theta_H\frac{1+\cos 2\varphi_H}{2}\right)\sin\omega t \\ + \frac{1}{4}LS_{ANE}KI_0^2\sin\theta_H\sin\varphi_H\sin^2\omega t, \quad (2)$$

where $\hat{x}$ is a unit vector along x-axis. After neglecting the effect of OMR, and substituting $\theta_H = 90°$ into Equation (2), the first harmonic component of $V_{arm-x}$ is:

$$V_{arm-x}^\omega = \frac{R_0' I_0}{2} + \frac{R_{AMR}}{4}I_0 \cos 2\varphi_H, \quad (3)$$

where $R_0' = R_0 + \frac{R_{AMR}}{2}$ and $R_{AMR} = \frac{L\Delta\rho_{AMR}}{A}$. The corresponding first harmonic resistance of arm-x ($R_{arm-x}^\omega$) is:

$$R_{arm-x}^\omega = R_0' + \frac{R_{AMR}}{2}\cos 2\varphi_H. \quad (4)$$

By following the same derivations on arm-y, we can obtain the first harmonic resistance of arm-y ($R^{\omega}_{arm-y}$) as:

$$R^{\omega}_{arm-y} = R'_0 - \frac{R_{AMR}}{2}\cos 2\varphi_H, \qquad (5)$$

Since $R_{AMR}$ is much smaller than $R_0$, we may write $R'_0 \approx R_0$. In the ideal case, when the four arms are identical, the first harmonic voltage of the bridge is given by

$$V^{\omega}_1 = \frac{I_0}{2}\left(R^{\omega}_{arm-x} - R^{\omega}_{arm-y}\right) = V_{AMR}\cos 2\varphi_H, \qquad (6)$$

where $V_{AMR} = \frac{I_0 R_{AMR}}{2}$ is the amplitude of $V^{\omega}_1$. To account for the non-ideality of the device, we may write $V^{\omega}_1$ as

$$V^{\omega}_1 = V_{AMR}\cos 2\varphi_H + C_{01}. \qquad (7)$$

Here, $C_{01}$ is the offset caused by misalignment or higher order angle-dependent terms due to inhomogeneous anisotropy in the sensing arms. The offset can be eliminated in the initial calibration process by subtracting half of the sum of the maximum and minimum values from $V^{\omega}_1$. After which the signal can be normalized by dividing it with the amplitude, $V_{AMR}$.

Similarly, by substituting the expressions of $\nabla T$, $\vec{H}_{ext}$, $\vec{J}_x$, $\hat{m}$ into Equation (1), ignoring OMR and OHE, the transverse voltage on arm-x ($V_2$) can be expressed as:

$$\begin{aligned}V_2 = W\vec{E}\cdot\hat{y} = &-\frac{WS_{ANE}KI_0^2}{8}\sin\theta_H\cos\varphi_H \\ &+ \left(\frac{W\Delta\rho_{AMR}I_0}{4A}\sin^2\theta_H\sin 2\varphi_H + \frac{W\rho_{AHE}I_0}{2A}\cos\theta_H\right)\sin\omega t \\ &+ \frac{WS_{ANE}KI_0^2}{8}\sin\theta_H\cos\varphi_H\cos 2\omega t,\end{aligned} \qquad (8)$$

where $W$ is the width of each arm, $\hat{y}$ is a unit vector along y-axis. Here, $\frac{W\Delta\rho_{AMR}I_0}{4A}$ term corresponds to the contribution of planar Hall effect (PHE). When $\theta_H = 90°$, $V^{2\omega}_2$ is:

$$V^{2\omega}_2 = \frac{WS_{ANE}KI_0^2}{8}\cos\varphi_H = V_{ANE2}\cos\varphi_H, \qquad (9)$$

where $V_{ANE2}$ is the amplitude of $V^{2\omega}_2$. Using the same analysis method on arm-y, $V^{2\omega}_3$ is obtained as

$$V^{2\omega}_3 = V_{ANE3}\sin\varphi_H, \qquad (10)$$

where $V_{ANE3}$ is the amplitude of $V^{2\omega}_3$. Considering the offsets, $V^{2\omega}_2$ and $V^{2\omega}_3$ may be written as:

$$V^{2\omega}_2 = V_{ANE2}\cos\varphi_H + C_{02}, \qquad (11)$$

$$V^{2\omega}_3 = V_{ANE3}\sin\varphi_H + C_{03}, \qquad (12)$$

where $C_{02}$ and $C_{03}$ are the transverse offset terms for the two arms, respectively. In practical case, $C_{02}$ and $C_{03}$ can be ignored as they are very small.

*3.2. Measured angle dependence of harmonic signals*

Figure 2a shows the relationship between $\varphi_H$ and $V^{\omega}_1$, obtained from a Wheatstone bridge consisting of CoFeB (6 nm)/MgO (2 nm)/Ta (1.5 nm), from bottom to top. The ac current applied has an amplitude of 15 mA and frequency of 115 Hz. The in-plane field applied is 500 Oe, which is sufficient to saturate the magnetization into field direction. The offset of $V^{\omega}_1$ has been subtracted from the raw data. The blue circle represents the measurement results, while the solid line depicts the fitting curves obtained using Equation (7). For this sample, the amplitude of $V^{\omega}_1$ is 21.5 mV and the offset is 14.9 mV. Figure 2b shows that the error term exhibits a $\cos 4\varphi_H$ dependence, which is presumably caused by the induced anisotropy introduced by the magnetic field applied during deposition, though its magnitude is very small, around 0.017 mV [23,24]. The overall fit error stands at 1.05%, indicating that the output curve exhibits high consistency with the cosine double-angle dependence.

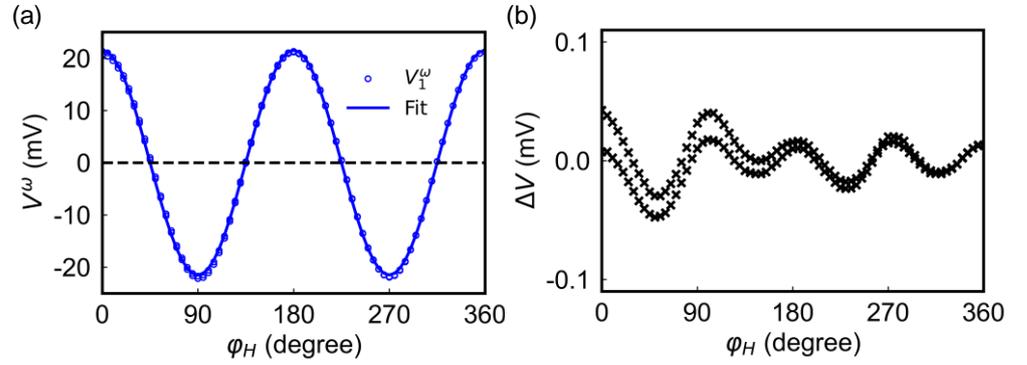

**Figure 2.** (**a**) Measured harmonic voltage $V_1^\omega$ (open circle), as a function of $\varphi_H$ and its fitting curve (solid line). The offset has been subtracted from the raw data. (**b**) Error distribution of $V_1^\omega$ as a function of $\varphi_H$.

Figure 3a shows the relationship between $\varphi_H$ and $V_2^{2\omega}$, $V_3^{2\omega}$, respectively, acquired under same conditions as that of $V_1^\omega$. The solid green and orange lines represent the fitting curves obtained using Equations (11) and (12), respectively. The angle dependence of $V_2^{2\omega}$ and $V_3^{2\omega}$ closely aligns with the fitting curves, indicating high consistency with the cosine and sine angle dependence, respectively. The amplitudes of $V_2^{2\omega}$ and $V_3^{2\omega}$ are nearly the same at 10.6 µV. The error in the ANE signal is attributed to the third-order harmonic error introduced by the Oersted field [19,25], with a magnitude of 0.12 µV, 0.14 µV for $V_2^{2\omega}$, $V_3^{2\omega}$, respectively and random error terms, as shown in Figure 3b. The fit errors for $V_2^{2\omega}$, $V_3^{2\omega}$ are 1.49%, 1.89%, respectively, indicating low noise of the measured signals.

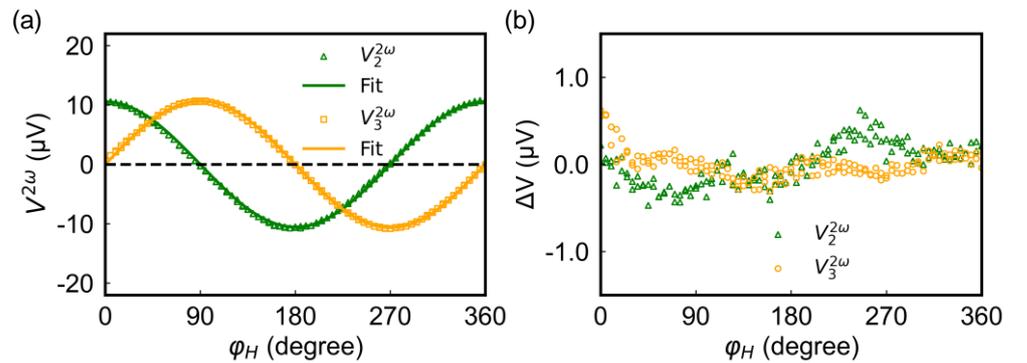

**Figure 3.** (**a**) Measured harmonic voltages $V_2^{2\omega}$ (open triangle) and $V_3^{2\omega}$ (open square) as a function of $\varphi_H$ and their fitting curves (solid line). (**b**) Error distribution of $V_2^{2\omega}$ and $V_3^{2\omega}$ as a function of $\varphi_H$.

*3.3. Angle calculation from harmonic signals*

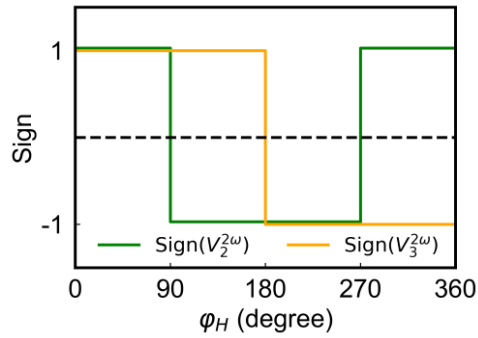

**Figure 4.** Sign of $V_2^{2\omega}$, $V_3^{2\omega}$.

We now proceed to the calculation of angles using the harmonic signals. There are two possible ways to calculate the angles using the combination of first and second harmonic signals. The method 1 involves the use of the first harmonic signal to calculate the angles in each quadrant and then using the sign of second harmonic signal to determine actual angles within 0° - 360°. As illustrated in Figure 4, the sign combination of $V_2^{2\omega}$ and $V_3^{2\omega}$ is unique in each quadrant: both are positive for 1st quadrant and negative for 3rd quadrant, and one is positive and the other is negative in the 2nd and 4th quadrant, respectively. The combination of arccosine values calculated from $V_1^{\omega}$ and the sign of $V_2^{2\omega}$ and $V_3^{2\omega}$ allows one to determine the angles within 0° - 360°. The specific calculation steps are as follows:

i) Calculate $\varphi'$ from Equation (7): $\varphi' = \frac{1}{2}\text{acos}\left(\frac{V_1^{\omega}-C_{01}}{V_{\text{AMR}}}\right)$, $\varphi' \in (0°, 90°)$;

ii) Determine the actual angle ($\varphi$) according to the sign of $V_2^{2\omega}$ and $V_3^{2\omega}$

$$\varphi = \begin{cases} \varphi' & V_2^{2\omega} > 0 \text{ and } V_3^{2\omega} > 0 \\ 180° - \varphi' & V_2^{2\omega} > 0 \text{ and } V_3^{2\omega} < 0 \\ 180° + \varphi' & V_2^{2\omega} < 0 \text{ and } V_3^{2\omega} < 0 \\ 360° - \varphi' & V_2^{2\omega} < 0 \text{ and } V_3^{2\omega} > 0 \end{cases}. \tag{13}$$

The angle calculated from this method is, in general, quite accurate, except for the region near the maximum or minimum of $\cos 2\varphi$, due to the relatively slow change in these regions. Therefore, the method 2 is to calculate the angle directly from $V_2^{2\omega}$ and $V_3^{2\omega}$ as follows:

$$\varphi = \text{atan2}\left(-\frac{V_3^{2\omega}}{V_{\text{ANE3}}}, -\frac{V_2^{2\omega}}{V_{\text{ANE2}}}\right) + \pi. \tag{14}$$

As is with the case of the first method, $\varphi$ calculated from $V_2^{2\omega}$ and $V_3^{2\omega}$ contains large error near the maximum and minimum values of $\cos \varphi_H$ and $\sin \varphi_H$. Therefore, we combine the two methods by setting proper boundary values to divide them for being used in different angle regions. By varying the boundary values between 0.8 and 0.99, we calculated the mean and maximum angle errors. The results indicated that when the boundary value is set between 0.86 and 0.9, both the mean and maximum angle errors reach their minimum values. Considering the larger amplitude of the first harmonic signal, we choose a larger value, 0.9, as the boundary value for distinguishing between the two calculation methods.

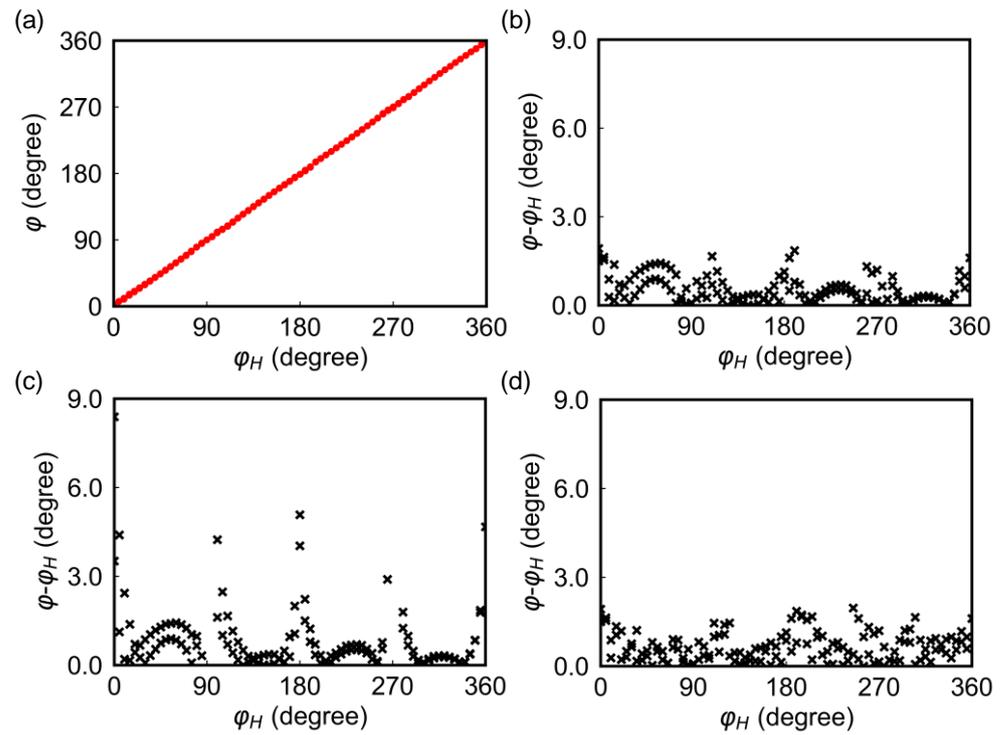

**Figure 5.** (**a**), (**b**) Angle φ and angle error distribution as a function of $\varphi_H$ using combined method, respectively. (**c**), (**d**) Angle error distribution using method 1 and method 2, respectively.

Figure 5 a,b show φ and angle errors obtained from the combined method. Figure 5a demonstrates a notable consistency between the actual magnetic field angle and the calculated angles from the measurement data. The angular distributions calculated independently using the method 1 and method 2 are depicted in Figure 5 c,d, with mean angular errors of 0.85° and 0.67°, respectively, and maximum angle errors of 8.39° and 1.92°, respectively. The combined method effectively reduces both mean and maximum angle errors, as shown in Figure 5b. The angle error is within the range of 0.01° to 1.92° with a mean error of 0.54°, which is comparable to commercial devices with multiple sensors…

*3.4. Performance optimization*

3.4.1. Effect of FM layer thickness

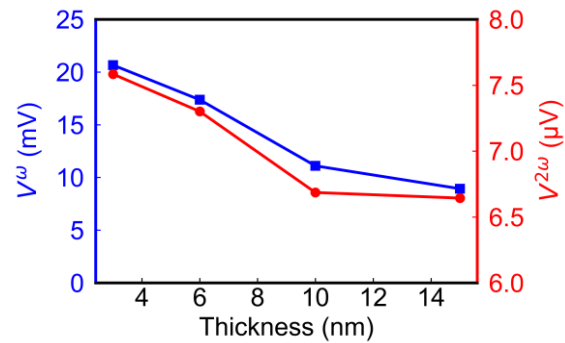

**Figure 6.** Relationship between the thickness of CoFeB and the amplitude of first (red round) and second harmonic (blue square) signals.

Figure 6 shows the amplitudes of $V_1^\omega$ (square) and $V_2^{2\omega}$ (circle) as a function of CoFeB thickness obtained with a fixed magnetic field strength of 500 Oe and same dissipation power of 190 mW (by varying the current). The measurements were performed on samples of four different thicknesses: 3, 6, 10, and 15 nm. It is observed that as the thickness increases, the amplitudes of both $V_1^\omega$ and $V_2^{2\omega}$ decrease. Consequently, thinner samples under the same power consumption condition results in a higher output voltage. However, it's important to note that at a thickness of 3 nm, the sample is susceptible to over-heating due to high resistance. Therefore, a CoFeB thickness of 6 nm is chosen in this study.

3.3.2. Effects of current amplitude

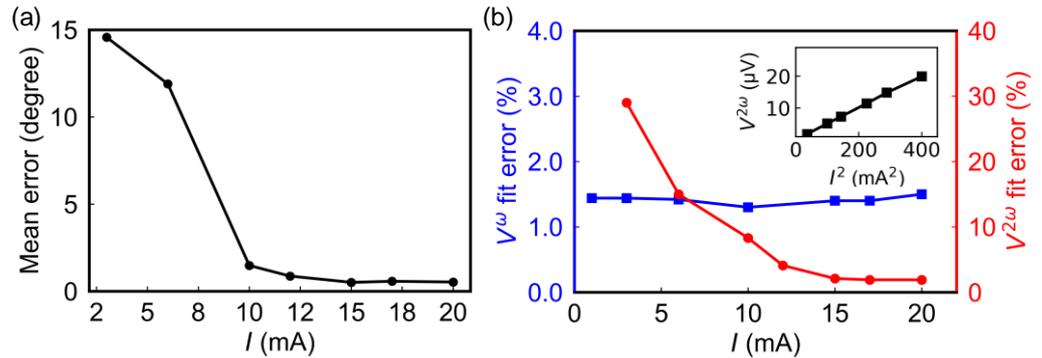

**Figure 7.** (**a**) Mean angle error calculated with different current amplitudes using a sensor with 6 nm CoFeB. (**b**) The relationship between current and fit errors of the first and second harmonic signals. Insert shows the proportional relationship between the amplitude of second harmonic signal and current square.

After determining the thickness of the FM layer, we investigated the effect of current amplitude on the sensor output signals and angle errors. Figure 7a illustrates the mean angle error calculated with different current amplitudes using a 6 nm thick sensor. At low currents, the mean angle error is large, reaching approximately 14°. As the current amplitude increases to 15 mA and above, the mean angle error stabilizes at around 0.51°. The reason is that the fit error of the output signals significantly affects the accuracy of angle calculation, as shown in Figure 7b. The blue and red dashed lines depict the relationship between the fit errors of the first and second harmonic signals with the current, respectively. At low currents, the fit error of the second harmonic signal is more pronounced, reaching up to 30%, primarily due to small temperature gradient. When the current exceeds 15 mA, the second harmonic signal's fit error reaches a minimum. The inset in Figure 7b reflects the proportional relationship between the second harmonic signal and the square of the current amplitude, which is due to $\nabla T \propto I^2$, as mentioned above. On the other hand, the fit error of the first harmonic signal remains relatively stable within measurement current range. This stability is attributed to the effective compensation provided by the Wheatstone bridge structure, mitigating changes in resistance on different arms of the Wheatstone bridge caused by temperature rises. Taking energy efficiency into consideration, we set the current amplitude at the inflection point of the second harmonic signal's fit error, i.e., 15 mA, as the input for the sensor.

*3.5. Magnetic field and temperature dependence*

Harmonic voltages $V_1^\omega$, $V_2^{2\omega}$ and $V_3^{2\omega}$ as a function of $\varphi_H$ under a wide range of magnetic fields are shown in Figure 8a-c. The amplitude of the first and second harmonic signals are between 21.13 – 21.98 mV and 10.65 – 11.16 μV, respectively within the 100 – 10,000 Oe range. The solid line in Figure 8d illustrates the output curve of a commercial Hall effect angle sensor, TMAG5170 [26] at 1500 Oe. When the external magnetic field is larger than 1,000 Oe, which is the maximum detection field of the TMAG5170 sensor, it becomes ineffective, failing to produce an ideal sinusoidal or cosine curve as the dashed lines shown in Figure 8d. However, for our sensor, when the field is larger than 300 Oe, the curves overall remain the same amplitude, indicating its weak dependence on the strength of applied field. This can be advantageous in practical applications with a diverse field range.

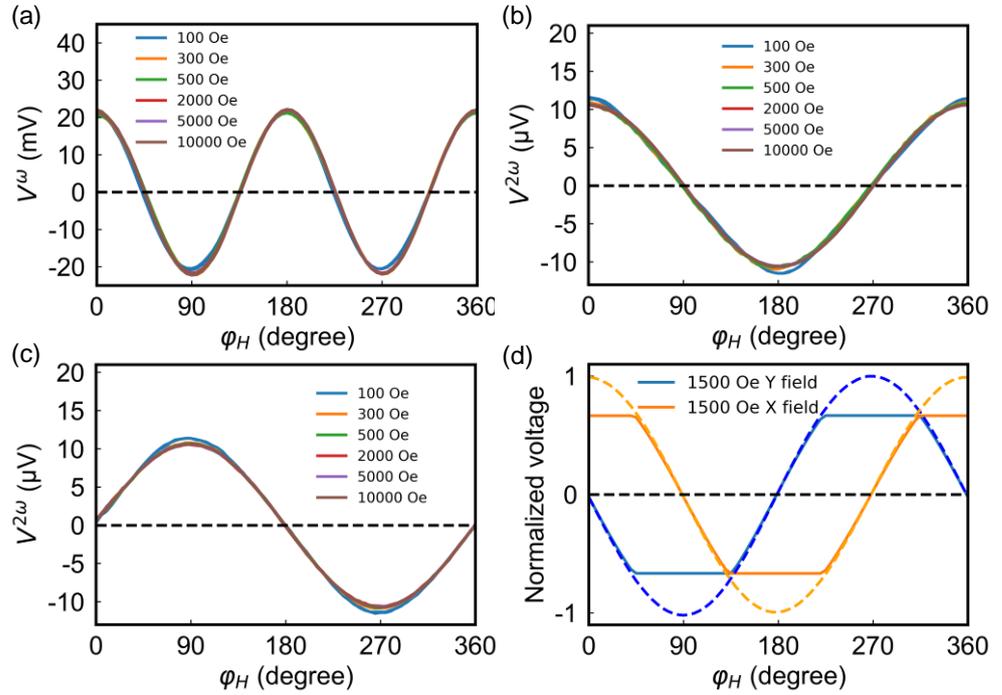

**Figure 8.** (**a**), (**b**), (**c**) Harmonic voltages $V_1^{\omega}$, $V_2^{2\omega}$ and $V_3^{2\omega}$ as a function of $\varphi_H$ under a wide range of magnetic fields. (**d**) Normalized voltage curves of TMAG5170 under 1500 Oe (solid lines). The dashed lines are ideal output curves when the field is within the detection range.

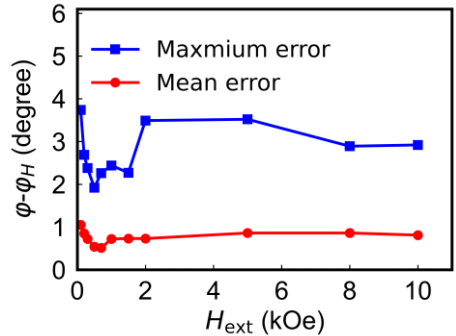

**Figure 9.** Mean and the maximum angle errors as a function of external field ranging from 100 to 10,000 Oe at 298 K.

We use the same combined calculation method to calculate the angles under different external field in the range of 100 - 10000 Oe. As shown in Figure 9, overall, the mean and maximum angle errors exhibit very small variation at different magnetic fields. The mean angle error (red circle) fluctuates between 0.51° and 1.05° within 100 Oe to 10000 Oe range, reaching its minimum around 0.51° between 500 Oe and 700 Oe, while the maximum angle error (blue square) ranges from 1.9° to 3.7° within the field range.

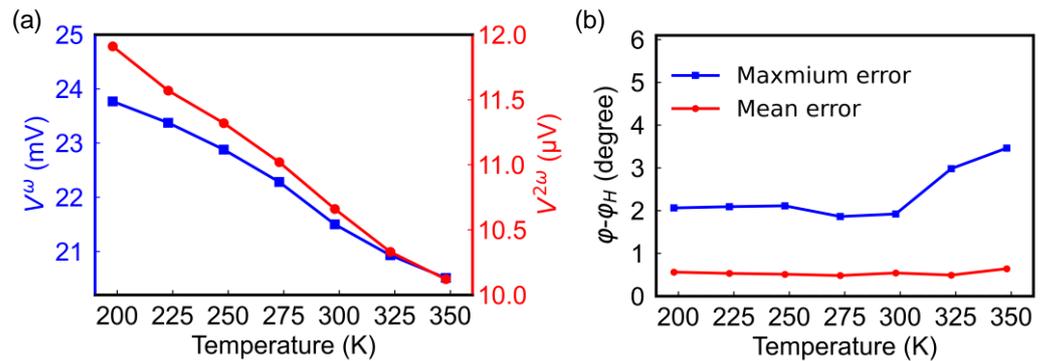

**Figure 10.** (**a**) Mean and the maximum angle error within temperature range of 198 – 348 K at fixed 500 Oe field. (**b**) First-order and second-order signal amplitudes as a function of temperature.

Finally, we also investigate the temperature effect. As the temperature rises, the scattering-dependent extrinsic contributions to AMR effect decrease [27], which results in the monotonic decrease of the first harmonic signal amplitude, ranging from 20.51 to 23.77 mV with a change of 15.9%. Additionally, the temperature elevation in the ambient condition leads to a reduction in the temperature gradient, causing the second harmonic signal amplitude to decrease monotonically, as shown in Figure 10a. Therefore, when using the sensor at different temperatures, recalibration at each temperature is necessary to ensure accuracy. Under the temperature range of 198 to 348 K at 500 Oe, the mean angle error remains stable at 0.49° – 0.64°, while the maximum angle error is ranging from 1.9° to 3.4°, as shown in Figure 10b.

## 4. Discussion

Compared to full 360° commercial magnetic position sensors and SOT-based angle sensors as shown in Table 1 [10,13,16,17,25], the sensor presented in this work exhibits several notable advantages, particularly its structural simplicity. Utilizing a single Wheatstone bridge structure, our sensor distinguishes itself from other magnetic field sensors that often necessitate the combination of multiple devices. Another key advantage is the stability of the signal across a wide magnetic field range (100 – 10,000 Oe). The absence of field-dependent components in the signal suggests that, in principle, the sensor can operate in a very large field range. This unique property makes the sensor particularly well-suited for diverse magnetic field applications. However, we also notice that at elevated temperatures, especially when surpassing 100 °C, establishing a stable temperature gradient becomes challenging. The mean and maximum angle errors may increase at elevated temperature, which is currently the limitation of this type of sensor.

**Table 1.** Comparison of full 360° magnetic position sensors.

| Model | Number of devices | Dynamic range (Oe) | Mean error (degree) [1] | Temperature range (°C) |
|---|---|---|---|---|
| TMR/GMR sensors [10] | 2 | 300 - 500 | 0.7 | -40 - 150 |
| Hall effect sensors [25] | 2 | 20 – 1,000 | 0.4 | -40 - 150 |
| AMR sensors with GMR sensors [13] | 3 | 200 - 600 | 0.1 | -40 - 125 |
| SOT vector magnetometer [16] | 1 | 0 - 50 | 1.1 | - |
| SOT-based sensor [17] | 1 | 500 – 2,000 | 0.4 | - |
| AMR/ANE sensor in this work | 1 | 100 – 10,000 | 0.5 | -80 - 80 |

[1]: Here only lists the mean errors within their optimal magnetic field ranges.

## 5. Conclusions

In conclusion, we have designed an alternative type of magnetic position sensor based on the ANE and AMR. It consists of a single Wheatstone bridge and can measure in-plane magnetic field angles within the full range of 0° - 360°. The combination of AMR and ANE signals reduces both mean and maximum angle errors. The prototype device exhibits a mean angle error of 0.51° at field range of 500 - 700 Oe at room temperature, and further performance improvement and applicability expansion can be achieved through material and device geometry optimization. Considering its simple structure and suitability for wide magnetic field ranges, our design may present a cost-effective approach for magnetic position sensing.

**Author Contributions:** Y.H.W. conceived the idea and supervised the project. J.Q.W. designed and fabricated the device and performed the measurements. J.Q.W. and Y.H.W. analyzed the data. H. X. performed the experimental work of the project at the initial stage. J.Q.W. and Y.H.W. wrote the manuscript.

**Funding:** This work is supported by the Advanced Research and Technology Innovation Centre (ARTIC), the National University of Singapore under Grant A-0005947-23-00 and NUS GAP Funding (RIE2025) under Grant A-8000656-00-00.

**Institutional Review Board Statement:** Not applicable.

**Informed Consent Statement:** Not applicable.

**Data Availability Statement:** Data are contained within the article.

**Conflicts of Interest:** The authors declare no conflict of interest.